\documentclass[floatfix,twocolumn,prb,usletter]{revtex4}

\usepackage{epsfig}

\begin{document}

\title{Exchange biasing of single-domain Ni nanoparticles spontaneously grown
in an antiferromagnetic MnO matrix}

\author{Daniel P. Shoemaker} \email{dshoe@mrl.ucsb.edu}
\author{Madeleine Grossman} 
\author{Ram Seshadri} 

\affiliation{Materials Department and Materials Research Laboratory\\
University of California, Santa Barbara, CA, 93106, USA}

\date{\today}

\begin{abstract}

Exchange biased composites of ferromagnetic single-domain Ni nanoparticles 
embedded within large grains of MnO have been prepared by reduction of
Ni$_x$Mn$_{1-x}$O$_4$ phases in flowing hydrogen. The Ni precipitates
are 15-30\,nm in extent, and the majority are completely encased within
the MnO matrix.  The manner in which the Ni nanoparticles are spontaneously
formed imparts a high ferromagnetic--antiferromagnetic
interface/volume ratio, which results in substantial exchange bias
effects.  Exchange bias fields of up to 100\,Oe are observed, in cases
where the starting Ni content $x$ in the precursor Ni$_x$Mn$_{1-x}$O$_4$ phase
is small. For particles of approximately the same size,
the exchange bias leads to significant hardening of the magnetization, with 
the coercive field scaling nearly linearly with the exchange bias field.

\end{abstract}

\maketitle 

\section{Introduction} 

Exchange anisotropy, or exchange bias, is an interfacial phenomenon between 
ferromagnetic and antiferromagnetic domains which results in the shifting and 
broadening of magnetic hysteresis loops.  Exchange bias is believed to result 
from the interaction of ferromagnetic (FM) spins with uncompensated
antiferromagnetic (AFM) spins at the FM/AFM 
interface.\cite{blamire_uncomp,kuch_uncomp,nogues_schuller}  
Since its discovery in partially oxidized Co/CoO nanoparticles by Meiklejohn 
and Bean,\cite{mbean} exchange bias has been observed and engineered in
core-shell nanoparticles,\cite{coreshell} thin films,\cite{thinfilms} and 
granular composites.\cite{nanostructure_review} These
architectures are utilized because a high proportion of FM spins must be
interfacial in order for the AFM switching behavior to appreciably
affect the FM coercivity.  While they achieve a high interface/volume
ratio, core-shell nanoparticles and thin film architectures
do not result in large quantities of exchange-biased material.
As an alternative, novel methods of processing exchange biased systems have
been explored, including coevaporation,\cite{coevap} mechanical
milling,\cite{sort_milling} and spontaneous phase 
separation.\cite{separation_alloy} 

Initial reports from Sort \textit{et al.}\cite{sort} have demonstrated 
hydrogen reduction of Fe$_{0.2}$Cr$_{1.8}$O$_3$ to produce metal/oxide 
composites. Different transition metals reduce sequentially, resulting in
nanosized Fe particles within micron sized Cr$_2$O$_3$ grains. 
Interaction between the $\sim$10\,nm Fe precipitates and the bulk
Cr$_2$O$_3$ provides exchange bias shifts of 10\,Oe.  Reduction
kinetics of the system CoCr$_2$O$_4$--Co$_3$O$_4$ have been reported by
Bracconi and Dufour\cite{bracconi}, and Kumar and Mandal\cite{kumar} have
produced Co/Cr$_2$O$_3$ composites directly from nitrate precursors.  
Recently, Toberer \textit{et al.}\cite{toberer_advmat05} have demonstrated 
that remarkable microstructures with aligned porosity can be
observed when the reduction product shares a common oxygen sublattice
with the precursor. 

Here we report on hydrogen reduction of the system Ni$_x$Mn$_{3-x}$O$_4$ 
to form Ni/MnO composites with striking microstructures associated
with substantial exchange biasing.
The Ni particles exhibit bulk saturation magnetization values, and
exchange bias is observed below the N\'eel temperature of MnO at $T_N$ =  
119\,K.  Surface and interior particle size analysis reveals
that this system produces Ni nanoparticles on the order of 15\,nm to 30\,nm. 
Size-dependent exchange bias phenomena are manifested in trends between
the Ni content of the precursor spinel and the exchange and coercive
fields of the reduced composite.

\section{Experimental}

Single-phase ceramic monoliths were prepared
by solid-state reactions of oxalates, similar to that of
Wickham.\cite{wickham} Oxalates are versatile precursors for 
mixed metal oxides, and have found extensive use in recent years
to produce substituted binary\cite{risbud_co-zno,lawes_co-mn-zno} and 
ternary\cite{toberer_chemmat05,toberer_advmat05,toberer_chemmat06} compounds. 

Stoichiometric amounts of
nickel acetate and manganese acetate [Ni(CH$_2$COOH)$_2$$\cdot$4H$_2$O and
Mn(CH$_2$COOH)$_2$$\cdot$4H$_2$O, Aldrich 99\,\%] were added to a solution
containing one equivalent of glacial acetic acid.  Excess oxalic acid
monohydrate [H$_2$(C$_2$O$_4$)$\cdot$H$_2$O, Fisher 99.9\,\%] was mixed in a
separate solution and both were stirred at $90^{\circ}$C.  Addition of
the oxalic acid to the dissolved acetates results in coprecipitation of
very fine, single-phase nickel-manganese oxalates in which the metals are 
mixed on the atomic scale. The oxalate powders, 
Ni$_x$Mn$_{3-x}$(C$_2$O$_4$)$_3$$\cdot$2H$_2$O,
were washed with deionized water and dried at $90^{\circ}$C,
calcined in alumina boats in
air at temperatures ranging from 780$^\circ$ to $1200^\circ$C for 10\,h,
then quenched into water to prevent conversion to $\alpha$-Mn$_2$O$_3$ or
to NiMnO$_3$.  The resulting single-phase Ni$_x$Mn$_{3-x}$O$_4$ spinel
powder was pressed into pellets at 100\,MPa and sintered at
$1325^{\circ}$C for 24\,h, then annealed at the previous calcination
temperature and water quenched.

Reductions were performed in 
alumina boats in a tube furnace under 
5\,\% H$_2$/N$_2$ with a flow rate of approximately 30\,sccm. Once the gas 
mixture had equilibrated, the specimens, as pellets, were heated at 
$2^{\circ}$C/min to 650$^\circ$C, 700$^\circ$C, or 725$^\circ$C, held for 
2\,h, then cooled at $10^\circ$C/min to room temperature. Reduced samples 
were verified to be Ni/MnO by x-ray diffraction (XRD, Philips X'Pert with 
Cu$K_{\alpha}$ radiation) and Rietveld refinement using the \textsc{xnd}
code.\cite{xnd} Composites were characterized by
thermogravimetic analysis (TGA, Cahn TG-2141), scanning electron
microscopy (SEM, FEI Sirion XL40), focused ion beam milling and
microscopy (FIB, FEI DB235), and SQUID magnetometry (Quantum Design MPMS
5XL). 

\section{Results and Discussion}

\begin{figure}
\centering\epsfig{file=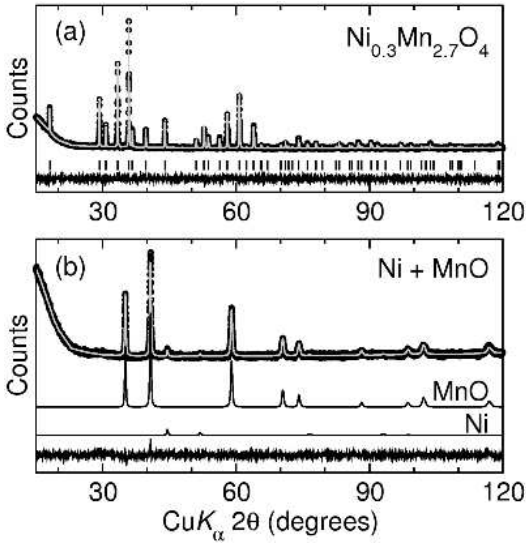,width=7cm}\\
\caption{X-ray diffraction Rietveld refinements of 
(a) Ni$_{0.3}$Mn$_{2.7}$O$_4$ single-phase tetragonal spinel (hausmannite)
precursor, and (b) the \textit{fcc}-Ni/rock-salt MnO composite produced by 
reduction of the above spinel in 5\,\%H$_2$/N$_2$.}
\label{rietveld10}
\end{figure}

The calcining of the single-phase Ni/Mn oxalates, according to the phase 
diagram presented by Wickham,\cite{wickham} results in single phase  
spinel-related compounds that are not all cubic. Wickham\cite{wickham}
has reported that in their high-temperature state, bulk samples 
of Ni$_x$Mn$_{3-x}$O$_4$ with $x$ between 0.15 and 1.00 are cubic spinels 
before decomposing into NiMnO$_3$ and $\alpha$-Mn$_2$O$_3$ in the temperature
range of 705$^\circ$ to 1000$^\circ$C.  Upon
water quenching, samples prepared with $x < 1$ and fired at
$\geq1000^{\circ}$C are observed to distort from the high-temperature 
cubic spinel reported by Wickham into single-phase hausmannite-type tetragonal
spinels in space group $I4_1/amd$. Slow-cooling, air-quenching, or quenching 
in flowing nitrogen are insufficient to prevent decomposition of the solid
solution. 

Rietveld refinement of the room-temperature XRD pattern for the
water-quenched compound Ni$_{0.30}$Mn$_{2.7}$O$_4$ is shown in 
Fig.\,\ref{rietveld10}(a).  Only peaks for the hausmannite-type solid solution
are evident; this is a requirement for the final reduced composite to be 
homogeneous in terms of the distribution of Ni precipitates.
The refinement assumes a ``normal" spinel, where Ni$^{2+}$ and Mn$^{2+}$ 
occupy the 4b tetrahedral sites. Mn$^{3+}$ in the 8c octahedral 
sites causes a cooperative Jahn-Teller distortion which leads to a loss of 
cubic symmetry.\cite{goodenough_jt}  An accurate determination of the cation
distribution may be obtained by neutron diffraction and has been investigated 
by Larson \textit{et al.}\cite{cation_dist}  When sintered at 1325$^{\circ}$C, 
samples with $x$ near 1 partially decompose into mixtures of NiO and
Ni$_{1-\delta}$Mn$_{2+\delta}$O$_4$ as described by Wickham,\cite{wickham} 
but subsequent annealing at 800$^{\circ}$C for 72\,h
ensures the formation of a single-phase tetragonal spinel.  Dense pellets
and micron-sized powder are both suitable precursors for hydrogen
reduction because the dimensions of the precipitates and pores are
orders of magnitude smaller than the grain size in either case. 
Adequately high oxygen mobility at the reduction temperature allows the 
reaction to permeate the sample regardless of any lack of preexisting porosity.

\begin{figure}
\centering\epsfig{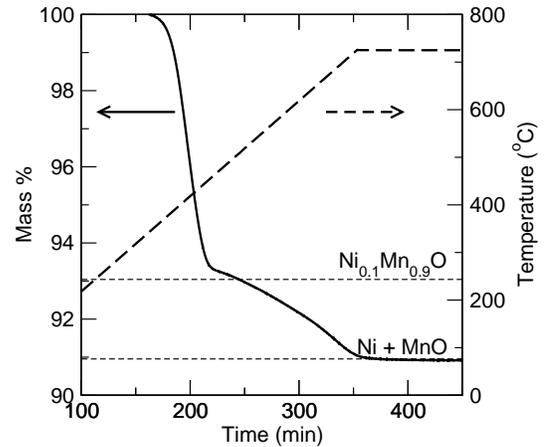}\\
\caption{TGA of a Ni$_{0.3}$Mn$_{2.7}$O$_4$ sample shows that reduction 
proceeds by an initial reaction to rocksalt Ni$_{0.1}$Mn$_{0.9}$O solid 
solution, followed by a reduction of Ni$^{2+}$ into metallic Ni.} 
\label{tga}
\end{figure}

In all cases, TGA analysis confirms the total amount
of nickel precipitated (and thus the stoichiometry of the precursor
spinel) during hydrogen reduction.  A TGA weight loss curve for
Ni$_{0.3}$Mn$_{2.7}$O$_4$ is shown in Fig.\,\ref{tga}.  The
weight loss curve reveals that the single-phase spinel first reduces to
a rocksalt (Ni$_{0.1}$,Mn$_{0.9}$)O solid solution, followed by
precipitation of metallic Ni. This progression is verified by
the fact that incompletely reduced samples display an MnO lattice
parameter that is smaller than the theoretical value, due to Ni substitution.
X-ray diffraction Rietveld refinement of the final composite
product obtained after reduction in 5\% H$_2$/N$_2$ indicate only
rocksalt MnO and face-centered cubic Ni [Fig.\,\ref{rietveld10}(b)].

\begin{figure}
\centering\epsfig{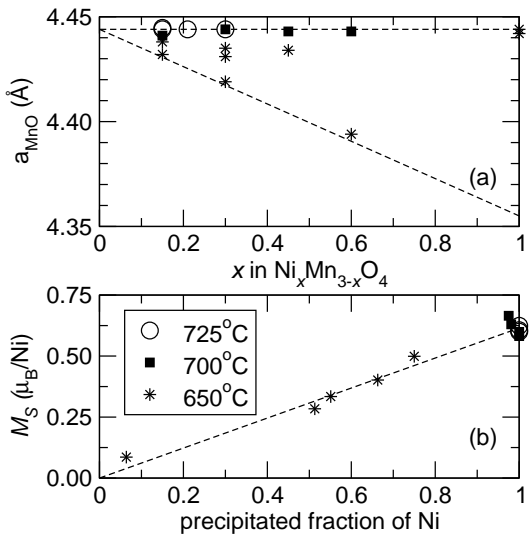}\\
\caption{(a) Lattice parameter of MnO obtained by Rietveld refinement 
of samples after hydrogen reduction at varying temperatures. The diagonal 
dotted line is the calculated lattice parameter of a (Ni,Mn)O solid solution, 
while the top line represents the desired conversion to pure MnO.
(b) Magnetic saturation of reduced Ni/MnO samples increases linearly 
with the completeness of Ni reduction as determined by $a_{\rm{MnO}}$
from XRD. All magnetic data concerning coercivity or 
exchange bias was measured from samples with complete Ni reduction and 
thus $M_S \approx 0.6 \mu_B$/Ni.}
\label{reducheck}
\end{figure}

High-spin Mn$^{2+}$ in octahedral coordination has an ionic radius of 
0.83\,\AA\/ in contrast to octahedral Ni$^{2+}$ which has a radius of only 
0.69\,\AA.\cite{shannon-prewitt} Consequently, when Ni$^{2+}$ enters product
MnO lattice, there is significant shrinkage of the cell parameter, which can 
be used to estimate the degree of conversion of the starting phases into pure 
Ni/MnO. The MnO lattice parameter obtained from Rietveld refinement is plotted
in Fig.\,\ref{reducheck}(a) as a function of the Ni content in the single-phase
hausmannite/spinel precursor. The cell parameter of pure MnO, 4.444\,\AA\,\ is 
also indicated as a horizontal dashed line. It is seen that for 
small substitution of Ni ($x$ in the starting phases) the reduction 
temperature must be increased from 650$^\circ$C to 725$^{\circ}$C to ensure 
complete reduction and avoid the rock-salt (Ni,Mn)O solid solution.  
Depression of the required reduction temperature of Ni$_x$Mn$_{3-x}$O$_4$ as 
$x$ deviates from Mn$_3$O$_4$ is a consequence of the higher ionization 
energy of Ni$^{2+}$.  In other words, more energy is released by reduction 
of Ni$^{2+}$ ions than of Mn$^{2+}$, so the reduction to metal occurs more 
readily when $x$ is larger. The greater ease of reduction of Ni over Mn
is suggested by the appropriate Ellingham diagram.\cite{ellingham} 
The saturation magnetization $M_S$ of the magnetic
Ni nanoparticle precipitates can be used in tandem with the values of 
$a_{\rm{MnO}}$ 
obtained from Rietveld refinement to determine the completeness of Ni 
reduction.  This is shown in Fig.\,\ref{reducheck}(b), where 
agreement is seen between the convergence of $a_{\rm{MnO}}$ and $M_S$ to their 
respective theoretical values of 4.444~\AA\, and 0.6\,$\mu_B$/Ni for a
completely reduced $x$Ni/MnO composite, regardless of $x$.

\begin{figure}
\centering\epsfig{file=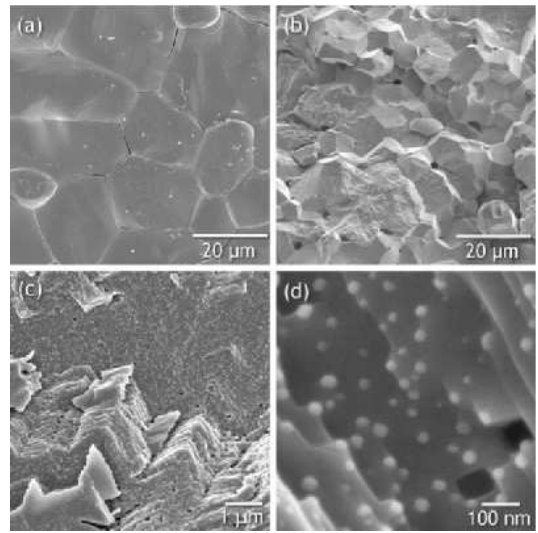,width=7cm}\\
\caption{Representative scanning electron microscope images: 
(a) The as-sintered surface of a dense pellet of Ni$_x$Mn$_{3-x}$O$_4$ with 
$x$ = 0.3. (b) The fracture surface of the material obtained from reducing
the sample in (a) at 725$^\circ$C for 2~h. (c) and (d) are the sample
in (b) shown at higher magnification. The highly porous and crystalline matrix
of MnO is seen in (c), and at higher magnification, small Ni particles with
sizes in the 30\,nm to 40\,nm range are seen as bright objects on a darker
background.}
\label{semprog}
\end{figure}

Hydrogen reduction of single-phase oxide monoliths can lead to striking 
hierarchically porous microstructures, which have been characterized by 
Toberer \textit{et al.}\cite{toberer_advmat05,toberer_chemmat06,toberer_chemmat07}
At first glance, low-magnification SEM micrographs of Ni$_x$Mn$_{3-x}$O$_4$
precursor spinels and Ni/MnO reduced samples [Fig.\,\ref{semprog}(a)
and Fig.~\ref{semprog}(b), respectively] appear nearly identical.  However,
higher magnification [Fig.\,\ref{semprog}(c) and (d)]
reveals that reduced composites contain aligned pores in rock-salt MnO
covered with Ni metal nanoprecipitates.  It has been previously 
suggested \cite{toberer_advmat05,toberer_chemmat07} that the shared
oxide sublattice of spinel and rocksalt allows the transformation from
one to the other to take place without reconstruction. 
Porosity is introduced during the spinel to rocksalt transformation 
while leaving the oxygen framework largely intact. The associated volume loss
gives rise to a pore structure that can be regarded as negative crystals --
voids in crystals that possess the same facets as the crystals themselves do. 

Although the pores are as small as 20\,nm, the pore and surface edges are 
aligned at right angles over the entire breadth of the 20\,$\mu$m grains.  This
long-range alignment implies that the MnO grains are in fact single
crystals with the same orientation and extent as the pre-reduction
spinel grains.\cite{toberer_advmat05,toberer_chemmat06}  
Increasing the reduction temperatures should lead to densification 
and closing of the pores in the MnO monolith.
However, in the interest of maintaining
small Ni nanoparticles (and thus a high interface/volume ratio), and
because the majority of nanoparticles are completely encased in MnO even
in porous samples, reduction was performed at the lowest temperature
that allowed complete Ni precipitation.

\begin{figure}
\centering\epsfig{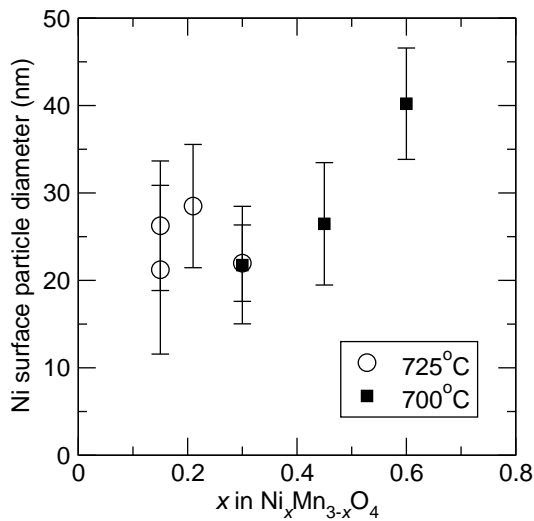}\\
\caption{The mean diameters of the Ni particles on the surface of the MnO matrix
as a function of the initial Ni content $x$ in Ni$_x$Mn$_{3-x}$O$_4$.
Error bars indicate one standard deviation in the particle diameter. Typically
at least 30 distinct particle's were counted in preparing the distributions. 
It is seen that most in most samples, the sizes are somewhat independent of $x$
and are clustered around 30\,nm.}
\label{size_nifrac}
\end{figure}

If we assume that for the different values of $x$, the number of nuclei are 
the same, and that increasing $x$ only affects the growth (\textit{ie.} the 
diameter) of the particles, then we would expect only a weak dependence 
(changing as $x{^\frac13}$) of the particle diameter rate on $x$. 
If we assume that increasing $x$ also increases the number of Ni nuclei 
upon reduction, then average particle diameter would show an even weaker 
dependence on $x$.
We have analyzed the Ni particles in the SEM images of the surfaces of the
monoliths by using the program \textsc{imageJ}\cite{imagej} to prepare 
histograms of particle size distributions. These are plotted in 
Fig.\,\ref{size_nifrac} for the different
monoliths. It is seen that mean particle diameters 
range from $\sim$15\,nm to 35\,nm, but there is no clear trend in size,
at least until a nickel content of $x = 0.60$ is reached. 

\begin{figure}
\centering\epsfig{file=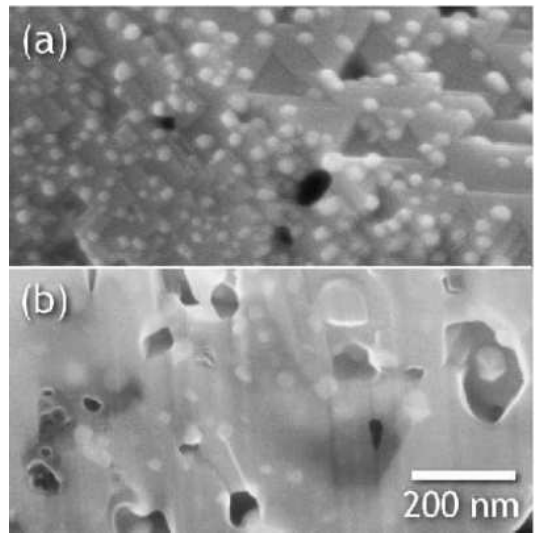,width=7cm}\\
\caption{(a) Fracture surface of a reduced sample (700$^\circ$C, 2\,h)
of Ni$_x$Mn$_{3-x}$O$_4$ with $x$ = 0.45, and (b) the surface of a 
FIB-cut sample of the product formed on reduction (700$^\circ$C, 2\,h)
of Ni$_x$Mn$_{3-x}$O$_4$ with $x$ = 0.60.
The two images are displayed at the same magnification. It is seen in (a) 
that different faces of the underlying MnO
seem to nucleate different particle sizes of \textit{fcc}-Ni. In (b),
it is seen that the Ni particles are found within the MnO matrix as well, and
not simply on the surface.}
\label{fib}
\end{figure}

Indeed, in the different monoliths, a clearer correlation is found 
for Ni particle size with the specific crystallographic face of MnO
upon which it grows, rather than the starting $x$ value.
It is evident in Fig.\,\ref{fib}(a) that for a $x = 0.45$
specimen, regions can be found which exhibit a wide variety of surface
particle sizes and spacings depending on the nucleation environment. 
The coherent pore structure introduced by reduction produces square or
triangular facets seen in Fig.\,\ref{fib}(a) which correspond to
exposed \{100\} or \{111\} faces.

Cross-sections of reduced grains produced by FIB milling, shown in 
Fig.\,\ref{fib}, reveal that the bulk
MnO contains Ni nanoparticles of similar dimensions as those on the surface. 
Porosity is still prevalent in the bulk of the monoliths as it is in the
images of the monolith surface. This is necessary to accommodate the volume loss
of the structure while retaining the size and alignment of the MnO
grains.  By a comparison of lattice parameters, and assuming no
sintering during reduction, the fraction of intragranular porosity
produced by the conversion of Ni$_x$Mn$_{3-x}$O$_4$ to $x$Ni/MnO
increases linearly from 16\% when $x=0$ to 39\% when $x=0.6$, which is
in rough agreement with observations of the intragranular pore volume in
FIB-milled samples.  Most Ni nanoparticles observed in
cross-section [Fig.\ \ref{fib}(b) are completely encased within the MnO 
matrix.  Based upon the observed surface density of Ni particles and assuming
100\,nm diameter pores, it can be determined
that the observed surface Ni particles only constitute about 20\% of the volume of
Ni that must be precipitated.  Therefore, we estimate that
approximately 80\% of the Ni grains are encased within the MnO matrix.

\begin{figure}
\centering\epsfig{file=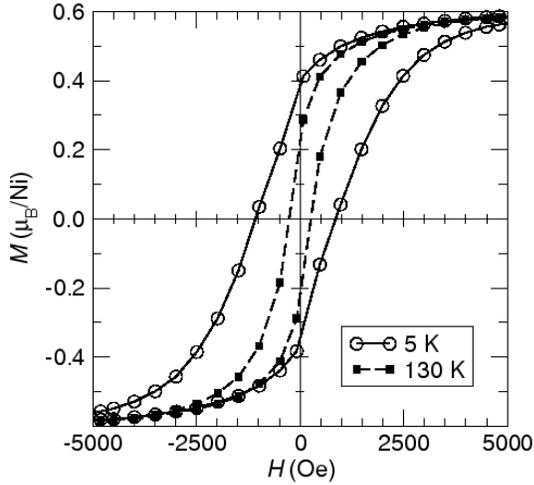,width=7cm}\\
\caption{Magnetization $M$ as a function of magnetic field $H$ 
for an Ni/MnO composite with $x = 0.3$ above $T_{N}$ (dashed) and field-cooled 
under a 50\,KOe field to 5\,K (solid). 
Exchange bias leads to a broadening  (associated with the coercivity $H_C$)
and shift (associated with the exchange field $H_E$) of the 
field-cooled loop at 5\,K.}
\label{hys}
\end{figure}

\begin{figure}
\centering\epsfig{file=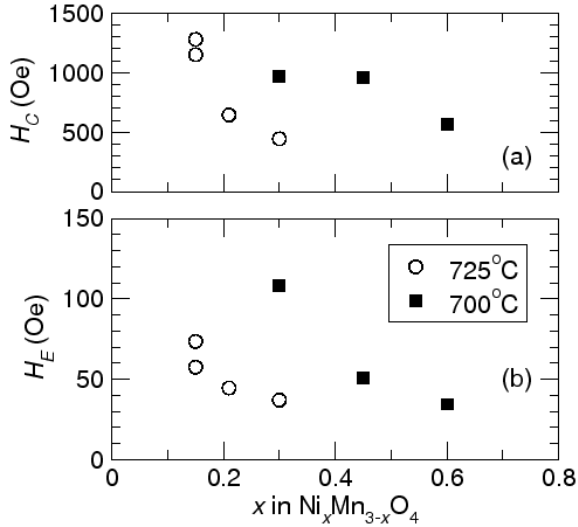,height=7cm}\\
\caption{(a) Coercive field $H_C$ and (b) Exchange-bias field $H_E$ as a 
function of the initial Ni content $x$ in the reduced Ni/MnO composites. 
The coercivity $H_C$ of the Ni nanoparticles decreases with Ni content for 
each firing temperature in response to the increased fraction of interfacial 
spins as Ni content is decreased. Exchange bias effects become less pronounced 
as well, with increasing Ni content in the oxide precursor.} 
\label{mag_nifrac}
\end{figure}

Magnetic hysteresis loops for an $x = 0.3$ sample (Fig.\ \ref{hys}) display
at 5\,K, a loop shift $H_E$ characteristic of exchange
biased systems.  Above the N\'{e}el temperature of MnO, $T_N$ = 119\,K,
the hysteresis loop is centered about $H = 0$.  After field cooling at
$H$ = 50\,kOe, the coercive field is broadened
and shifted $H_E$ = 100\,Oe in opposition of the cooling field direction.
The exchange behavior can be influenced by many
factors, including Ni particle size, the amount and orientation of the
FM-AFM interface, temperature, and the cooling field.\cite{nogues_schuller}
We anticipate that in the size regime studied here (near 20\,nm) the Ni
nanoparticles are single-domain magnets and that the coercivity below
the blocking temperature should not show a strong size-dependence.\cite{cullity}
Fig.\ \ref{mag_nifrac}(a) shows that as the
nickel content $x$ increases, $H_C$ decreases for samples reduced at
either 700$^{\circ}$C or 725$^{\circ}$C. 
At both reduction temperatures, the highest $H_C$ is found for the smallest 
$x$, and the smallest $H_C$ is found for the largest $x$.  

Additionally, the decrease in $H_C$ for $x$ = 0.3 samples reduced at 
725$^{\circ}$C as opposed to 700$^{\circ}$C implies increased coalescence 
of Ni particles as the temperature increases. We therefore anticipate that 
the increased coercivity as size is decreased arises from the same
interfacial coupling that results in the increased exchange bias.

In exchange biased nanostructures of spherical FM particles in an AFM
matrix, the strength of the exchange field $H_E$ has been suggested to
vary as

\[ H_E = \frac{6 E_A}{M_S d_{\rm{FM}}} \] 

\noindent where $E_A$ is the interfacial coupling energy per unit area, 
$M_S$ is the saturation magnetization of the FM, and $d_{\rm{FM}}$ is the 
diameter of the FM particle.\cite{nanostructure_review} Assuming this model
to be correct, we anticipate that the exchange field $H_E$ should decrease 
with in increasing ferromagnetic particle size. If, with increasing $x$ in our 
systems, Ni particle particle size indeed increases, then our results are 
broadly consistent with this model.

\begin{figure}
\centering\epsfig{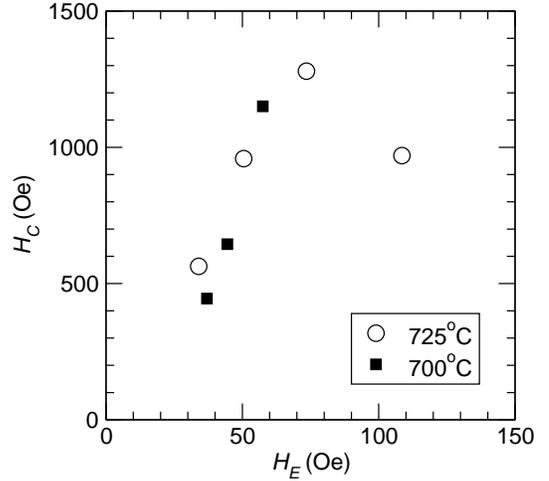}\\
\caption{Coercive field as a function of exchange field at 5\,K for the
different Ni/MnO composites. For most of the composite systems, $H_C$ varies 
nearly linearly with $H_E$.}
\label{hc_xb}
\end{figure}

In Fig.\,\ref{hc_xb} we plot the 5\,K coercivity as a function of the exchange
field for the different systems measured, data for which are displayed in 
Fig.\,\ref{mag_nifrac}. We see that the coercivity varies nearly linearly with
the exchange field, with the exception of one outlier. G\"okemeijer 
\textit{et al.}\cite{PhysRevB.63.174422} have recently measured biasing of 
ferromagnets on different CoO surfaces and have concluded that on the 
uncompensated CoO surfaces, exchange biasing, and the associated shift of 
hysteresis is found, but on compensated CoO surfaces, the effect of the 
interface is simply to increase coercivity. The magnetic structure of MnO
is not simple\cite{goodwin:047209} and the architectures described here of
nearly spherical ferromagnetic particles embedded in an antiferromagnetic
host cannot be described in terms of simple interfaces. Given this, we suggest 
that perhaps both effects, of the uncompensated as well as the compensated
surfaces are playing a role, and the linear relation between coercivity and 
exchange is simply an indication of increasing interfacial area between the
two magnetic components. 

\section{Conclusions} 

We have demonstrated that hydrogen reduction of Ni$_x$Mn$_{3-x}$O$_4$
spinels produces Ni/MnO composites with significant interfacial area
between antiferromagnetic MnO and ferromagnetic Ni, and associated exchange 
bias. With increasing nickel content $x$, these effects decrease, presumably
because of a decrease in the relative proportion of interfacial spins
in the ferromagnet. Exchange bias effects at the FM--AFM interface lead to 
an increase in $H_C$ with decreasing Ni content, along with a $1/x$ dependence 
of $H_E$.  A nearly linear relationship is found between $H_C$ and $H_E$ in 
these systems.

\section{Acknowledgments}

This work was supported by the donors of the American Chemical Society 
Petroleum Research Fund, and the National Science Foundation through
a Career Award (DMR 0449354) to RS, and for the use of MRSEC facilities
(DMR 0520415). MG was supported by a RISE undergraduate fellowship.

\bibliographystyle{apsrev} \bibliography{MnO_Ni.bib}

\clearpage

\end{document}